\documentclass[prb,twocolumn,aps,noshowpacs]{revtex4}
\usepackage{graphicx,epsfig}
\usepackage{bm}
\usepackage{amssymb} 
\usepackage{bbold}

\begin{document}
\newcommand{\s}{\scriptscriptstyle}

\title{Two-photon absorption in a two-level system enabled by noise}

\author{V. V. Mkhitaryan$^1$, C. Boehme$^1$, J. M. Lupton$^{1,2}$,  and M. E. Raikh$^1$}

\affiliation{ $^1$ Department of Physics and
Astronomy, University of Utah, Salt Lake City, UT 84112 \\
$^2$ Institut f{\"u}r Experimentelle und Angewandte Physik, Universit{\"a}t Regensburg, Universit{\"a}tsstr. 31, 93053 Regensburg, Germany}

\begin{abstract}
We address the textbook problem
of dynamics of a spin placed in a dc magnetic
field and subjected to an ac drive. If the drive
is polarized in the plane perpendicular to the
dc field, the drive photons are resonantly absorbed when the spacing between the Zeeman levels is close to the photon energy. This is the
only resonance when the drive is circularly polarized.
For linearly polarized drive, additional
resonances corresponding to  absorption
of three, five, and multiple odd numbers of photons is possible.
Interaction with the environment causes the broadening of the absorption lines.
We demonstrate that the interaction with environment
{\em enables} the forbidden two-photon absorption.
We adopt a model of the environment in the form of random telegraph noise produced by a single fluctuator.
As a result of the synchronous time fluctuations
of different components of the random field,
the shape of the two-photon absorption line is non-Lorentzian and depends dramatically on the drive
amplitude. This shape is a monotonic curve at strong
drive, while, at weak drive, it develops a two-peak structure reminiscent of an induced transparency on resonance.

\end{abstract}

\maketitle

\section{Introduction}
It has been known for a long time\cite{Autler1955,Shirley1965,book}
that a spin placed in a magnetic field, ${\bm B}={\bf z}_0B_0$,
and subjected to a linear ac drive, ${\bm B}_1={\bf x}_0B_1\cos \omega t$,
exhibits Rabi oscillations when the driving frequency, $\omega$,
is close to
\begin{equation}
\label{omegaP}
\omega_{2p+1}=\frac{B_0}{2p+1},
\end{equation}
where $p$ is integer. The frequency of $(2p+1)$-photon Rabi oscillations is, within a factor,
given by\cite{Shirley1965} $\frac{B_1^{2p+1}}{B_0^{2p}}$. Coupling to the environment\cite{GrifoniReview} causes the decay of the
Rabi oscillations, i.e. dephasing. In the frequency domain, two split Rabi $\delta$-peaks in the
absorption spectrum acquire a finite width and, upon further increasing of coupling to the environment, gradually merge into a single peak.

For some time now, \cite{q1,q2,q3,q4,q5,q6,q7,review,Galperin}
Rabi oscillations can be realized on individual
two-level  systems rather than on ensembles.
In these realizations, the role of $\uparrow$ and $\downarrow$
states of a spin is played by different charge- or flux-states of a
superconducting qubit.

Absence of even-photon absorption peaks follows
from the Floquet description of the driven two-level
systems. The argument presented in
Ref.~\onlinecite{Autler1955} is general and goes as follows.

The system relating the amplitudes
of $\uparrow$ and $\downarrow$ projections of spin reads

\begin{eqnarray}
\label{system} &&i{\dot C}_{\s \frac{1}{2}}=\frac{B_0}{2}C_{\s
\frac{1}{2}}
+B_1\cos \omega t~ C_{\s -\frac{1}{2}},\\
&&i{\dot C}_{-\s \frac{1}{2}}=-\frac{B_0}{2}C_{-\s \frac{1}{2}}
+B_1\cos \omega t~ C_{\s \frac{1}{2}}.\nonumber
\end{eqnarray}

Searching for solutions in the form
\begin{equation}
\label{Floquet}
C_{\s \frac{1}{2}}=e^{i\lambda t}\sum\limits_{-\infty}^{\infty}\alpha_n e^{in\omega t},~C_{-\s \frac{1}{2}}=e^{i\lambda t}\sum\limits_{-\infty}^{\infty}\beta_n e^{in\omega t},
\end{equation}
where $\lambda$ is the Floquet exponent,
reduces the system Eq.~(\ref{system}) to the infinite set of coupled equations
\begin{eqnarray}
\label{infinite}
&&\left[\left(\lambda+n\omega\right)-\frac{B_0}{2}\right]
\alpha_n=
\frac{B_1}{2}\left(\beta_{n-1}+\beta_{n+1}\right),\\
&&\left[\left(\lambda+n\omega\right)+\frac{B_0}{2}\right] \beta_n=
\frac{B_1}{2}\left(\alpha_{n-1}+\alpha_{n+1}\right).
\end{eqnarray}
In the limit of a weak drive, $B_1\ll B_0$, a single-photon
resonance corresponds to $\lambda \approx -\frac{\omega}{2}$. For
this $\lambda$ the brackets in front of $\alpha_1$ and $\beta_0$
become small. To find the value of $\lambda$ with accuracy up to
the Bloch-Siegert shift, it is sufficient to truncate the system
neglecting all $\alpha_n$ and $\beta_n$ except for $\alpha_1$ and
$\beta_0$. We then obtain $\lambda=\pm \frac{1}{2}
\left[\left(\omega-B_0  \right)^2+B_1^2    \right]^{1/2}$, which
is the Rabi result.

A two-photon resonance, if it were allowed,
would correspond to $\lambda \ll B_0$ so that the brackets
in front of $\alpha_1$ and $\beta_{-1}$ become small.
However, it is seen from the system Eq.
(\ref{infinite}) that $\alpha_1$ and $\beta_{-1}$
are completely decoupled. Indeed, $\alpha_n$ with $n$
odd is coupled only to $\beta_n$ with $n$ even and vice versa.

\begin{figure}[h!]
\centerline{\includegraphics[width=80mm,angle=0,clip]{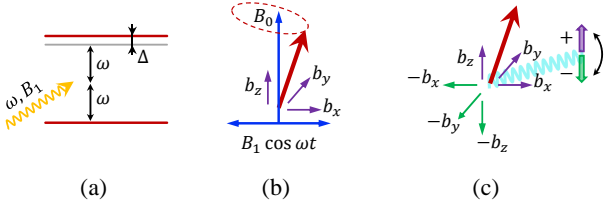}}
\caption{(Color online) (a) Schematic illustration of the resonant
two-photon absorption. The frequency, $\omega$, of drive with
amplitude $B_1\ll \omega$ is tuned to the condition
$\Delta=\left(B_0-2\omega\right) \ll \omega$, where $\Delta$ is
the detuning; (b) For polarization of drive, $B_1\cos\omega t$, in
the plane perpendicular to the field, $B_0$, two-photon absorption
is forbidden. The environment can be viewed as a random magnetic
field ${\bf b}(t)=\left(b_x(t), b_y(t), b_z(t)\right)$ fluctuating
with time. Coupling of the two-level system to the environment
enables the two-photon absorption; (c) We assume that the random
field, ${\bf b}(t)$, is produced by a fluctuator coupled to the
two-level system via the dipole-dipole interaction. Thus, the
vector ${\bf b}(t)$ assumes two positions, $ \pm\left( b_x, b_y,
b_z\right)$,  and switches between them at random moments. As a
consequence, the components of ${\bf b}(t)$ fluctuate {\em
in-phase}. } \label{fluctuator}
\end{figure}

The message of the present paper is that for a drive frequency,
$\omega$, close to $\frac{B_0}{2}$,  coupling to the environment
{\em enables} the resonant two-photon absorption.
We calculate the shape of the absorption peak
as a function of the drive amplitude and
the parameters of the environment.

To illustrate the effect, we choose the simplest model of the
environment illustrated in Fig.~\ref{fluctuator}. It represents a single
 fluctuator with a telegraph-noise dynamics and coupled to
 the spin by the dipole-dipole interaction.
We also demonstrate that for a general form of the environment,
the magnitude of the two-photon absorption depends crucially
on whether different  components of the random magnetic fields
are correlated or not.

We propose that this unexpected two-photon absorption, enabled by ambient noise, should be observable in devices made of organic semiconductors such as organic light-emitting diodes (OLEDs).

\section{Effective Hamiltonian}
Consider a single fluctuator shown in Fig.~\ref{fluctuator}.
By virtue of the dipole-dipole interaction,
this fluctuator produces a random magnetic
field ${\bf b}(t)=\big(b_x(t), b_y(t), b_z(t) \big)$
at the location of the spin, see Fig.~\ref{fluctuator}.
As a result, the terms $\pm \frac{B_0}{2}$
in Eq.~(\ref{system}) become $\pm \frac{B_0+b_z(t)}{2}$,
while the term $B_1\cos \omega t$ becomes $B_1\cos\omega t +b_x(t)\pm ib_y(t)$.

In order to incorporate the environment
into the Floquet description, one has to
assume that, instead of the common factor $e^{i\lambda t}$,
the coefficients $\alpha_n$ and $\beta_n$ are themselves time-dependent.
This generates the terms ${\dot \alpha}_n$ and ${\dot \beta}_n$
on the left-hand sides of the system Eq. (\ref{infinite}).
In fact, only the derivatives ${\dot \alpha}_1$ and ${\dot \beta}_{-1}$
should
be kept. This is because we assume that the dynamics of the
environment is slow, i.e. $\omega \tau \gg 1$, where $\tau$ is
the characteristic correlation time of the environment.
Under this condition, all the terms ${\dot \alpha}_n$
and ${\dot \beta}_n$ other than ${\dot \alpha}_1$, ${\dot \beta}_{-1}$
are added to much bigger terms $\left(n\omega+\frac{B_0}{2}\right)\alpha_n$ and $\left(n\omega-\frac{B_0}{2}\right)\beta_n$,
respectively.

Coupling of the amplitudes $\alpha_1$ and $\beta_{-1}$ is mediated by the
intermediate states with amplitudes  $\alpha_0$, $\alpha_{-1}$ and $\beta_0$,
 $\beta_1$. Keeping only these six amplitudes, we arrive at the following truncated system
\begin{eqnarray}
\label{truncated}
&&i{\dot \alpha}_1 -\frac{\Delta +b_z}{2}\alpha_1 =\frac{B_1}{2}\beta_0 +\frac{b_{-}}{2}\beta_1,\nonumber\\
&&-\frac{B_0}{2}\alpha_0=\frac{B_1}{2}\left(\beta_{-1}  +\beta_1\right)+\frac{b_{-}}{2}\beta_0,\nonumber\\
&&\frac{B_0}{2}\beta_0=\frac{B_1}{2}\left(\alpha_{-1}  +\alpha_1\right)+\frac{b_{+}}{2}\alpha_0,\\
&&-B_0\alpha_{-1}=\frac{B_1}{2}\beta_0+\frac{b_{-}}{2}
\beta_{-1},\nonumber\\
&&B_0\beta_{1}=\frac{B_1}{2}\alpha_0+\frac{b_{+}}{2}
\alpha_{1},\nonumber\\
&&i{\dot \beta}_{-1} +\frac{\Delta +b_z}{2}\beta_{-1}
=\frac{B_1}{2}\alpha_0 +\frac{b_{+}}{2}\alpha_{-1}.\nonumber
\end{eqnarray}
where $b_{\pm}=b_x\pm ib_y$ and
\begin{equation}
\Delta = B_0-2\omega
\end{equation}
is the detuning from the two-photon resonance.
In the coefficients in front of the non-resonant
amplitudes we have set $\omega=\frac{B_0}{2}$.

The strategy to analyze the system Eq. (\ref{truncated})
is to express the intermediate amplitudes,  $\alpha_0$, $\alpha_{-1}$, $\beta_0$,
and $\beta_1$ in terms of $\alpha_1$ and $\beta_{-1}$.
As a first step, we express $\alpha_{-1}$ and $\beta_1$
from the fourth and fifth equations and substitute them into the second
and third equations. This yields
\begin{eqnarray}
&&\label{23} -\left(B_0+\frac{B_1^2}{2B_0}\right)\alpha_0-
b_{-}\beta_0=
B_1\left(\frac{b_{+}}{2B_0}\alpha_1+\beta_{-1}\right), \nonumber\\
&&\left(B_0+\frac{B_1^2}{2B_0}\right)\beta_0
-b_{+}\alpha_0=B_1\left(\alpha_1-\frac{b_{-}}{2B_0}
\beta_{-1}\right).
\end{eqnarray}
Clearly, the term $\frac{B_1^2}{2B_0}$ is much smaller than $B_0$
and can be neglected. This term contributes to the Bloch-Siegert
shift of the two-photon resonance. Solving the system Eq.
(\ref{23}) in the leading order, we find
\begin{eqnarray}
\label{solution1}
&&\alpha_0=-\frac{B_1}{B_0}\left( \frac{b_{+}+2b_{-}}{2B_0}\alpha_1+\beta_{-1}\right),\nonumber \\
&&\beta_0=-\frac{B_1}{B_0}\left( \frac{2b_{+}
+b_{-}}{2B_0}\beta_{-1}-\alpha_1\right).
\end{eqnarray}
Next, we express $\beta_1$ and $\alpha_{-1}$ using Eq.
(\ref{solution1}) and substitute them into the right-hand sides of
the first and sixth equations of the system Eq.~(\ref{truncated}).
Upon doing so, we conclude that the equations for $\alpha_1$ and
$\beta_{-1}$ reduce to the pair of the Schr{\"o}dinger equations
generated by the effective Hamiltonian

\begin{equation}
\label{effective}
{\hat H}_{\s eff}=
\big(\Delta +b_z(t)\big)S_z+{\tilde b}_x(t)S_x,
\end{equation}
where the effective field ${\tilde b}_x(t)$ is defined as
\begin{equation}
\label{effectivebx} {\tilde b}_x(t)=\frac{2B_1^2}{B_0^2}b_x(t).
\end{equation}
Note that the $b_y$-component
of the random field drops out of the effective Hamiltonian.
Technically, this happens because ${\hat H}_{\s eff}$ contains
the combination $b_{+}+b_{-}=2b_x(t)$.

\section{Absorption spectrum}

\subsection{General expression}
In calculating the absorption coefficient from the
Hamiltonian Eq. (\ref{effective}) we take advantage of
the fact that ${\tilde b}_x$ is small, namely ${\tilde b}_x\tau \ll 1$.
This justifies using the cumulant expansion\cite{vanKampen}
to describe  the dynamics of
the disorder-averaged spin.
In particular, the spin-spin correlation function,
$\overline{ S_z(0)S_z(t)}$, has the form
\begin{eqnarray}
\label{spin-spin}
\overline{ S_z(0)S_z(t)} =\exp\Big[-\int\limits_0^tdt_1\int\limits_0^{t_1}dt_2 K(t_1-t_2)   \Big] &&\nonumber \\
=\exp\Big[-\int\limits_0^tdt_1\left(t-t_1\right)K(t_1) \Big],&&
\end{eqnarray}
where the correlator, $K(T)$, is defined as
\begin{equation}
\label{correlator}
K(T)=\Big\langle {\tilde b}_x(0){\tilde b}_x(T)
\cos\Big[\Delta T +\int\limits_{0}^{T}dt'b_z(t')\Big]\Big\rangle.
\end{equation}
and $\langle ...\rangle$ stands for the averaging over the
realizations of random time-dependent magnetic fields. Analytical
form of the correlator is found in the Appendix within the model
of the fluctuator, for which random magnetic fields $b_z(t)$ and
$b_x(t)$ fluctuate {\em in-phase} and assume two values $\pm b_x$
and $\pm b_z$. The durations of intervals between switching events
from $b_x,b_z$ to $-b_x,-b_z$ are distributed following the
Poisson distribution
\begin{equation}
p(t_i)=\frac{1}{\tau}\exp\left(-\frac{t_i}{\tau}\right),
\end{equation}
where $\tau$ is the average time between the successive switchings.

It is convenient to cast the result obtained in the Appendix in the form
\begin{equation}
\label{slow-fast}
K(T)=\frac{{\tilde b}_x^2\tau \cos(\Delta T)} {\big(1-b_z^2\tau^2\big)^{1/2}} \left[
\frac{1}{\tau_{f}}\exp\!\left(\!-\frac{T}{\tau_{f}}\right)
-\frac{1}{\tau_{s}}\exp\!\left(\!-\frac{T}{\tau_s}\right)\right],
\end{equation}
where $\tau_f$ and $\tau_s$ denote the fast and slow relaxation times defined as
\begin{equation}
\label{taufs}
\tau_f=\frac{\tau}{1+
\big(1-b_z^2\tau^2\big)^{1/2}},\quad \tau_s=\frac{\tau}{1-
\big(1-b_z^2\tau^2\big)^{1/2}}.
\end{equation}
It should be emphasized that  Eq. (\ref{slow-fast})
is derived for $b_x(t)$, $b_z(t)$
fluctuating {\em in-phase}. A dramatic consequence of these {\em in-phase} fluctuations is the sign minus
between the two exponential terms. As a result, $K(T)$ changes sign as a function of $T$.
If only $b_x(t)$ fluctuated, as is the case considered in Refs. \onlinecite{BergiTelegraph,GalperinTLS,Galperin},
the two terms in Eq. (\ref{slow-fast}) would {\em add rather than subtract}.
Below, we will see that this difference has a dramatic effect on the absorption lineshape

\begin{equation}
\label{spectrum}
I(\Delta)=2\int\limits_0^{\infty}dt \cos(\Delta t)\overline{S_z(0)S_z(t)}
\end{equation}
using the spin-spin correlation function Eq. (\ref{spin-spin}).

\subsection{Shape of the absorption spectrum}

Technically, the calculation is performed with the help of
the relation
\begin{eqnarray}
\label{long}
&&\int\limits_0^tdt_1(t-t_1)\cos\left(\Delta t_1\right) \exp\Big(-\frac{t_1}{\tau_{\s 0}}\Big)\\
&&=\frac{t\tau_{\s 0}}{1+\Delta^2\tau_{\s 0}^2}-\frac{\tau_{\s 0}^2
\left(1-\Delta^2\tau_{\s 0}^2\right)}{\left(1+\Delta^2\tau_{\s 0}^2\right)^2}\nonumber\\
&&+\tau_{\s 0}^2\Bigg[\frac{\left(1-\Delta^2\tau_{\s
0}^2\right)\cos\Delta t -2\Delta\tau_{\s 0}\sin\Delta t }
{\left(1+\Delta^2\tau_{\s 0}^2\right)^2}
\Bigg]\exp\Big(-\frac{t}{\tau_{\s 0}}\Big).\nonumber
\end{eqnarray}
At long times, the term linear in $t$ dominates the result. With
correlator Eq. (\ref{slow-fast}), this term assumes the form
\begin{eqnarray}
\label{leading}
&&\int\limits_0^tdt_1\left(t-t_1\right)K(t_1)\\
&&=\frac{{\tilde b}_x^2\tau
t}{\big(1-b_z^2\tau^2\big)^{1/2}}\left[\frac{1}{1+\Delta^2\tau_f^2}-\frac{1}{1+\Delta^2\tau_s^2}
\right]. \nonumber
\end{eqnarray}
Note that, for $\Delta=0$, the linear term vanishes. This is the
consequence of the in-phase fluctuations of $b_x(t)$ and $b_z(t)$.

Further calculations are performed for the most interesting regime  of the two-photon absorption,
the ``spectral narrowing''
regime,\cite{Anderson1953,Kubo1,Anderson1962} $b_z\tau \ll 1$. Under these conditions, the fast and slow
times differ strongly, namely, $\tau_s \gg \tau_f$.
Indeed, expanding Eq. (\ref{taufs}) we get
\begin{equation}
\label{asymptote}
\tau_f=\frac{\tau}{2},~~\tau_s=\frac{2}{b_z^2\tau}.
\end{equation}

At zero detuning, the exponent in Eq. (\ref{spin-spin})
is dominated by subleading terms in Eq. (\ref{long}). Thus,
instead of Eq. (\ref{leading}) one has
\begin{eqnarray}
\label{subleading}
&&\int\limits_0^tdt_1\left(t-t_1\right)K(t_1)\Big\vert_{\Delta=0} \\
&&\approx {\tilde b}_x^2\tau \left[\tau_s \left(1- e^{-\frac
t{\tau_s}}\right) -\tau_f \left(1- e^{-\frac
t{\tau_f}}\right)\right].\nonumber
\end{eqnarray}
Since $\tau_s\gg \tau_f$, the second term can be dropped. It then
follows from Eq. (\ref{subleading}) that the magnitude of
absorption is governed by the dimensionless parameter
\begin{equation}
\label{beta}
\beta ={\tilde b}_x^2 \tau \tau_s =2~\frac{{\tilde b}_x^2}{b_z^2},
\end{equation}
so that for $\beta \sim 1$, the characteristic time
in the integral Eq. (\ref{spectrum}) is $t \sim \tau_s$.
It also follows from Eq. (\ref{leading}) that for $\beta \sim 1$
the characteristic detuning is $\Delta \sim \tau_s^{-1}$. Indeed,
in terms of parameter $\beta$, the leading term Eq. (\ref{leading})
can be cast in the form

\begin{figure}[h!]
\centerline{\includegraphics[width=90mm,angle=0,clip]{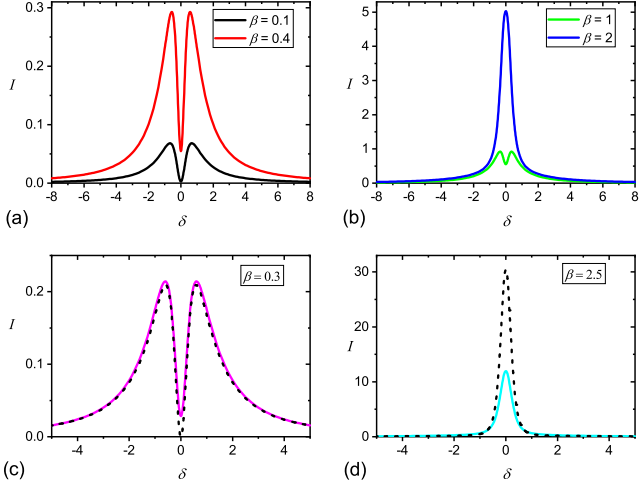}}
\caption{(Color online) The shapes of the two-photon absorption
spectra calculated numerically from Eq. (\ref{dimensionless1}) are
plotted for small (a) and large (b) values of the parameter
$\beta$ [Eq. (\ref{beta})]. For low beta values, a dip appears
around zero detuning. Upon the increase of $\beta \propto B_1^4$
the shape evolves from a two-peak structure to a single peak. Note
that the widths of the curves are only weakly dependent on
$\beta$. This dependency should be contrasted to the situation
when only the $b_x$ component fluctuates. In this situation, the
width grows linearly with $\beta$. In panels (c) and (d), the
numerical curves are shown together with the approximate
expressions Eq. (\ref{byparts}  ) and Eq. (\ref{zeroth-order}),
respectively (dotted lines). It is seen that Eq. (\ref{byparts})
describes the small-$\beta$ behavior fairly accurately, while Eq.
(\ref{zeroth-order}) captures the shape of the spectrum at large
$\beta$.} \label{fourplots}
\end{figure}

\begin{equation}
\label{leading1}
\frac{\beta \Delta^2\tau_s t}{1+\Delta^2\tau_s^2}.
\end{equation}
It is now convenient to rewrite the expression  Eq. (\ref{spectrum})
for the absorption spectrum by introducing the dimensionless variables
\begin{equation}
\label{dimensionless}
t_1=\frac{t}{\tau_s},~~\delta=\Delta\tau_s.
\end{equation}
In terms of these variables, Eq.~(\ref{spectrum}) assumes the form
\begin{eqnarray}
\label{dimensionless1}
&&I(\delta)=2\tau_s\int\limits_{0}^{\infty}dt_1 \cos \delta t_1
\exp\left(-\frac{\beta\delta^2t_1}{1+\delta^2}\right)\\
&&\exp\left\{\beta\frac{\left(1-\delta^2\right)
\left(1-e^{-t_1}\cos\delta t_1\right) +2e^{-t_1}\delta\sin\delta
t_1}{(1+\delta^2)^2} \right \}.\nonumber
\end{eqnarray}

Numerical plots of $I(\delta)$ for different $\beta$
are shown in Fig.~\ref{fourplots}. Our prime observation is that,
for strong drive, $I(\delta)$ is a monotonic curve, while
for a weak drive it develops a minimum at $\delta=0$.
We have traced this anomalous behavior to the fact
that the term $\cos\Delta t$ is present both as the
prefactor of Eq. (\ref{dimensionless}) and in the exponent.
We emphasize again that oscillations in the exponent
play a role only because the leading term Eq. (\ref{leading1})
is suppressed at small $\delta$,  which, in turn, is the consequence
of the in-phase fluctuations of
$b_x(t)$ and $b_z(t)$.
 Concerning the width of the curves, they narrow down upon
increasing drive.
Although it is impossible to evaluate the integral
Eq.~(\ref{dimensionless}) analytically,
we were able to capture the behavior of numerical curves
under certain assumptions.
If we neglect the oscillating terms in the exponent
completely, then the integral can be easily evaluated
yielding

\begin{equation}
\label{zeroth-order}
I(\delta)= 2\tau_s\beta \frac{1+\delta^2}
{\left(1+\delta^2\right)^2+\beta^2\delta^2}
\exp\left\{\beta\frac{1-\delta^2}{(1+\delta^2)^2}
\right\}.
\end{equation}
Not surprisingly, the above expression is a monotonic curve. For
$\beta \ll 1$, the shape, $I(\delta)$, is a Lorentzian:
$I(\delta)\propto \frac{1}{1+\delta^2}$, while for large $\beta$,
it is again a Lorentzian $I(\delta)\propto
\frac{1}{1+\beta^2\delta^2}$, but with a much smaller width.

In order to capture synchronous oscillations of the
prefactor and the exponent, we perform the integration
by parts in Eq.  (\ref{dimensionless}) and, subsequently,
again neglect the oscillations in the exponent.
The result again reduces to the elementary integrals and
has the form

\begin{eqnarray}
\label{byparts} &&I(\delta)= \tau_s\beta \left[
\frac{2(1+\delta^2)}
{\left(1+\delta^2\right)^2+\beta^2\delta^2} + \frac 1{1+\delta^2(1+\beta)}\right . \\
&&\left . - \frac{3(1+\delta^2)+\beta\delta^2}{4\delta^2
(1+\delta^2)^2+\big(1+\delta^2(1+\beta)\big)^2}
\right]\exp\left\{\beta\frac{1-\delta^2}{(1+\delta^2)^2} \right\}.
\nonumber
\end{eqnarray}
The first term in Eq. (\ref{byparts}) reproduces the previous
result Eq. (\ref{zeroth-order}). On the other hand, for $\Delta
=0$, Eq. (\ref{byparts}) yields zero. From Fig. \ref{fourplots} we
see that Eq. (\ref{zeroth-order}) describes the numerical results
at large $\beta$ (i.e. under the conditions of strong drive),
while Eq. (\ref{byparts}) captures a minimum at weak drive. In
other words, Eq. (\ref{zeroth-order}) completely neglects the
synchronous oscillations, while Eq.~(\ref{byparts}) overestimates
them. Roughly, one can argue that the true absorption spectrum
lies in between the analytical predictions Eq.
(\ref{zeroth-order}) and Eq. (\ref{byparts}). Recalling that the
parameter $\beta$ is proportional to the {\em fourth} power of
drive implies that the shape of the absorption spectrum is
extremely sensitive to the drive magnitude.

\subsection{Domain of applicability of the analytical results}

The form of the expressions Eqs. (\ref{zeroth-order}),
(\ref{byparts}) suggests that they do not apply in the domain
$\beta \gg 1$. Indeed, exponential growth of the absorption with
$\beta$ is unphysical. This prompts us to examine the validity of
the general expression Eq. (\ref{spin-spin}) for the spin-spin
correlator. We will see that the exponential growth is the
consequence of the in-phase fluctuations of $b_x(t)$ and $b_z(t)$.
For uncorrelated $b_x(t)$, $b_z(t)$ there is no unphysical growth
and the criterion ${\tilde b}_x\tau \ll 1$ is sufficient to
express the correlator in terms of the second cumulant.

The easiest way to derive  Eq. (\ref{spin-spin})
is to analyze the system of equations of motion
for the spin projections which follow from the
effective Hamiltonian Eq. (\ref{effective})
\begin{eqnarray}
\label{system}
&&\frac{\partial S_x}{\partial t}=-\left(\Delta +b_z(t)\right)S_y,\nonumber \\
&&\frac{\partial S_y}{\partial t}=-{\tilde b}_x(t)S_z+\left(\Delta +b_z(t)\right)S_x,\nonumber \\
&&\frac{\partial S_z}{\partial t}={\tilde b}_x(t)S_y.
\end{eqnarray}
Upon eliminating $S_x$ and $S_y$ and using the initial conditions
$S_x(0)=S_y(0)=0$, we arrive to the following closed integral equation for $S_z(t)$
\begin{eqnarray}
\label{closed}
&&S_z(t) =1 \\
&&-\!\int\limits_0^t\!\! dt_1\!\!\int\limits_0^{t_1}\!\! dt_2
{\tilde b}_x(t_2){\tilde b}_x(t_1)
\cos\Big[\Delta t_1+\!\int\limits_{t_2}^{t_1}\!dt'b_z(t')\Big]S_z(t_2).
\nonumber
\end{eqnarray}
This equation is exact and applies for an {\em arbitrary noise
realization}. It can be reduced to a closed integral equation for
$\overline{S_z(t)}$, the average over the noise realizations.
The form of the closed integral equation is the following \cite{vanKampen}
\begin{equation}
\label{avinteq}
\overline{S_z(t)} =1 -\!\int\limits_0^t\!\! dt_1\!\!\int\limits_0^{t_1}\!\! dt_2
K(t_1-t_2) \overline{S_z(t_2)},
\end{equation}
where $K(t)$ is the noise-averaged kernel in Eq. (\ref{closed})
given by Eq. (\ref{correlator}). Importantly, the reduction of Eq.
(\ref{closed}) to Eq. (\ref{avinteq}) is valid under the
assumption that the decay of $\overline{S_z(t)}$ is much slower
than the decay of $K(t)$. Solving Eq. (\ref{avinteq}) by
successive iterations reproduces the result Eq. (\ref{spin-spin}).
%

Now we have to check the basic assumption that $\overline
{S_z(t)}$ falls off slower than $K(t)$. If only $b_x(t)$
fluctuated, i.e. $b_z(t)=0$, then the time-scale for the change of
$K(t)$ would be $\tau$, [see Eq. (\ref{slow-fast})], while the
time-scale for the decay of $\overline {S_z(t)}$ would be
$\big({\tilde b}_x^2\tau\big)^{-1}$. This yields the conventional
condition ${\tilde b}_x\tau \ll 1$, which we assumed to be met.

For the in-phase fluctuations of $b_x(t)$ and $b_z(t)$, the
correlator falls off over much longer time $\tau_s\gg \tau$. On
the other hand, $\overline {S_z(t)}$ decays over time much longer
than $\big({\tilde b}_x^2\tau\big)^{-1}$. This decay time can be
estimated from Eq. (\ref{leading}) to be $t\sim \big({\tilde
b}_x^2\tau\big)^{-1}\left(\frac{1+\Delta^2\tau_s^2}
{\Delta^2\tau_s^2}\right)$.
Then the basic assumption reduces to the inequality
\begin{equation}
\label{condition}
\beta < \frac{1+\delta^2}{\delta^2}.
\end{equation}
This condition forbids large values of $\beta$, i.e. small values
of $b_z\ll \tilde{b}_x$. In the opposite limit of small $b_z$ the
absorption spectrum is Lorentizan with the width $\Delta ={\tilde
b}_x^2\tau$, i.e. it grows with drive as $B_1^4$.

\section{Discussion}

In this Section we discuss the experimental feasibility of the predicted
effect and put it into a general perspective.

\subsection{General remarks}

{\bf 1}. The phenomenon of two-photon absorption is widely used in
many different domains of science and engineering, primarily due
to its non-linear dependency on intensity which opens the
possibility of depth resolution in light-matter interaction, but
it is usually considered in the context of electronic dipole
transitions. Because of the angular momentum of the photon, the
parity of an electronic state changes after absorption of a photon
but remains the same under two-photon absorption. One-and
two-photon absorption therefore probe different electronic states
under electronic dipole transition. For magnetic dipole
transitions, in contrast, because of the change in spin quantum
number, the matrix element becomes non-vanishing between states of
the same parity, irrespective of whether one or two photons are
absorbed at once. Magnetic-dipole two-photon absorption therefore
offers a method to excite odd-parity states, which are spin
forbidden, and has been used to study para-excitons in
semiconductors and insulators \cite{N2}.


{\bf 2}. While our result of the absence of absorption at zero
detuning bears some superficial resemblance of conventional
electromagnetically induced transparency (EIT), analogues of which
can also be generated in magnetic-resonant systems, \cite{book,V2}
we stress that the origin is completely different. In EIT, quantum
interference occurs between different transition pathways of a
multilevel system, whereas  here, we consider solely the effect of
noise on two-level systems.

In fact, the origin of the dip at zero detuning is a direct
consequence of the in-phase telegraph-noise fluctuations of
$b_x(t)$ and $b_z(t)$. Indeed, upon setting $\Delta=0$ in the
effective Hamiltonian Eq. (\ref{effective}), it assumes the form
${\hat H}_{\s eff}=b_z(t)S_z+{\tilde b}_xS_x$. Note that the
eigenvalues of this Hamiltonian, $\pm \left[ b_z(t)^2+{\tilde
b}_x(t)^2\right]^{1/2}$, are {\em time-independent}. As $b_z$,
${\tilde b}_x$ change to $-b_z$, $-{\tilde b}_x$, two eigenvectors
remain the same while their eigenvalues interchange. As a result,
despite the randomness of the switching moments, $S_z(t)$
oscillates with time. Hence, there is no absorption.

\subsection{Relevance  to organic semiconductors}

{\bf 1}. In addition to superconducting qubits\cite{q1, q2, q3,
q4, q5, q6, q7, M2}, a natural setup to verify our predictions is
nuclear magnetic resonance, see e.g. Ref. \onlinecite{Glenn1},
although identification of individual spectral lines requires very
high sensitivity and selectivity which can be hard to achieve
given the limits imposed on sampling frequency by the small
nuclear gyromagnetic ratio. On the other hand, the two-photon
absorption predicted above is proportional to the fourth power of
the ratio $B_1/B_0$, and is generally very weak. Spin resonance in
electronic systems appears to offer a more promising avenue.
Indeed, a very high sensitivity of optical and electrical
detection of magnetic resonances has recently been demonstrated in
organic semiconductors, see Refs.~ \onlinecite{Lupton1, Lupton2,
Lupton3, Lupton4, Bayliss}. Crucially, instead of measuring
magnetic polarization, these experiments report on the permutation
symmetry of pairs of spins, and are therefore much more sensitive
than conventional measurements of magnetization. Also, organic
semiconductors are appealing because of the long spin lifetimes of
carriers. \cite{X2}

{\bf 2}. Experimental observation of the predicted effect
necessitates strong drive conditions where the carrier-wave
frequency $B_0$ is of order of the Rabi frequency $B_1$ and, in
addition, both frequencies are comparable to that of the noise.
Since MHz magnetic resonances are detectable electrically in
OLEDs, \cite{X3} such systems should offer an avenue for observing
forbidden two-photon
transitions.

{\bf 3}. In contrast to other well-defined spin systems such as
color centers in diamond, spin-1/2 paramagnetic species in organic
semiconductors do not experience any appreciable zero-field
splitting. As a consequence, magnetic resonance can even occur at
zero external field, mediated alone by the local hyperfine fields.
\cite{X3} Electrically detected magnetic resonance (EDMR) in OLEDs
bears some resemblance to magnetic-field effects and magnetic
resonance phenomena in radical-pair-based chemical reactions
\cite{X4, X5} but offers more flexibility in terms of choosing the
crucial ratio between carrier-wave frequency ($B_0$), Rabi
frequency ($B_1$) and the noise amplitude, which is determined by
the local hyperfine fields. These fields, in turn, can be tuned
chemically by controlling the isotopic ratio of hydrogen and
deuterium. \cite{X6} OLEDs offer the crucial advantage in EDMR
experiments that the active device area undergoing resonance can
be shrunk in size almost arbitrarily. This control is crucial not
only to enable the generation of high resonant driving field
amplitudes, \cite{X7} but also to ensure a high level of
homogeneity of the $B_0$ field. The latter aspect is crucial in
controlling the degree of static disorder of the Zeeman-split
two-level system with respect to the magnitude of dynamic
disorder, which enables the dipole-forbidden two-photon
transition.

{\bf 4}. Since an OLED operates by spin-dependent recombination of
electrons and holes, EDMR does not, strictly, probe a two-level
system, but two effectively degenerate two-level systems of the
Zeeman-split electron and hole spin levels. Under strong drive,
when the drive amplitude exceeds the variation in the expectation
value of the hyperfine fields, electron and hole resonance become
indistinguishable so that a new Dicke-type "superradiant" spin-1/2
resonant species emerges. \cite{X6} Since spin-orbit coupling and
resulting shifts of the gyromagnetic ratio are negligible at low
frequencies, \cite{X8} hyperfine coupling provides the only source
of inhomogeneous broadening of the resonance. Recently, clear
Zeeman resonances in EDMR were reported at frequencies as low as 5
MHz for a hydrogenated organic semiconductor; \cite{X3} for
deuterated materials, this limit is expected to be much lower. It
should therefore be straightforward to find a suitable parameter
space where the resonance frequency is of order the typical noise
frequency.


\section{Concluding remarks}

{\bf 1}. In spirit, the noise-induced absorption bares some
similarity to the hyperfine-induced excitation of the electron
spin resonance predicted and observed in Ref.
\onlinecite{Rashba+Marcus}. In simple terms, the idea of Ref.
\onlinecite{Rashba+Marcus} can be explained as follows. Suppose
that the driving field, $B_1$, is absent, but the quantum dot
confining an electron is ``shaken" in space by an ac electric
field with resonant frequency. As a result, hyperfine fields of
nuclei inside the dot also change with time with resonant
frequency and play the role of $B_1$. Since the hyperfine fields
constitute the environment, the resonance observed in Ref.
\onlinecite{Rashba+Marcus} can be viewed as environment-mediated.

In fact, the setup in experiment\cite{Rashba+Marcus} included two
quantum dots. Beating of electrons between the dots can be viewed
as pseudospin dynamics. The shaking of the dots then couples the
true electron spin dynamics to pseudospin dynamics.\cite{Rashba}


{\bf 2}. Description of the ac absorption within the Bloch
equations, \cite{Bloch} where the bath is modeled by the
spin-lattice relaxation time, $T_1$, does not capture the
two-photon resonance. In order to replicate a number of nontrivial
effects, a microscopic model of the environment must be specified.
Moreover, it is necessary to go beyond the Bloch-Redfield
description, as discussed in e.g. Ref.
\onlinecite{VagharshRotary}. For example, modeling of the
environment by a periodic modulation of the applied field, can
also allow two-photon absorption to occur. \cite{Glenn1}

{\bf 3}. While we have used the language of a spin driven by an ac
magnetic field, the results apply quite generally for any driven
two-level system. Usually, as, e.g., in Refs.
\onlinecite{Grifoni0,Grifoni,Burkard,Burkard1,x+z,x+z1,mph1,mphNori,mphFrench},
the effect of the environment on multiphoton absorption is studied
under the conditions where this absorption is allowed even in the
absence of environment. By contrast, our main message here is that
the environment itself can actually enable an absorption which is
otherwise forbidden.

{\bf 4}. Qualitative explanation of the absence of even-photon
resonances without an interaction with the bath is that linearly
polarized drive has matrix elements $\qquad \qquad |\downarrow
\rangle \rightarrow |\uparrow\rangle$ and $|\uparrow
\rangle\rightarrow |\downarrow \rangle$, but not  $|\downarrow
\rangle \rightarrow |\downarrow \rangle$ or $|\uparrow \rangle
\rightarrow |\uparrow \rangle$. Thus, each interaction with the
drive is accompanied by a spin-flip. Even number of interactions
with the drive necessarily returns the spin to the initial state,
thus forbidding the absorption of the even number of drive quanta.
As mentioned in Ref. \onlinecite{Shirley1965} and elaborated on in
Refs.  \onlinecite{Grifoni}, \onlinecite{x+z1}, a tilt of the dc
field away from the plane normal to the drive polarization results
in the matrix elements $|\downarrow \rangle \rightarrow
|\downarrow \rangle$ and $| \uparrow \rangle \rightarrow |
\uparrow \rangle$ becoming nonzero, thus allowing the  absorption
of an even number of drive quanta. In our study, the dc field is
normal to the drive, yet the fluctuating field produced by the
environment can be viewed as a time-dependent tilt of the dc
field.

{\bf 5}. A formal reason why in-phase fluctuations of $b_x(t)$ and
$b_z(t)$ result in an unconventional shape of the two-photon
absorption is the following. The argument of the cosine in the
correlator Eq. (\ref{correlator}) contains the integral $\int_0^T
dt b_z(t)$, and the prefactor has the form  ${\tilde
b}_x(0){\tilde b}_x(t)$. As a result of the in-phase fluctuations,
${\tilde b}_x(t)$ and $b_z(t)$ are proportional to each other.
Thus, the prefactor is proportional to the derivative of the
argument of the cosine, so that the correlator itself becomes
proportional to the {\em full time derivative of the oscillating
function}. The leading term Eq. (\ref{leading}) is proportional to
the integral of the correlator. Thus, it is not surprising that
this integral is zero for $\Delta=0$. We then conclude that for
environment of a spin representing not a single but an ensemble of
fluctuators, for which $b_x$ and $b_z$ are different but fluctuate
in-phase, the above property still holds. Thus, we would
intuitively expect the double-peaked  shape of the absorption
spectrum to emerge.

{\bf 6}. It is instructive to put our results into a more general
perspective of spectral
narrowing\cite{Anderson1953,Kubo1,Anderson1962}. The randomness of
$b_z$ emulates the spectral narrowing. However, if $b_x$
fluctuates in phase with $b_z$, the effect of spectral narrowing
is {\em undone}. The same effect can be reformulated in the
language of Franck-Condon physics. Within the Franck-Condon
picture, the optical matrix elements are renormalized due to
coupling of the electronic level to nuclear vibrations. If,
hypothetically, the transition between electronic levels was due
to the {\em same} vibrations, the Franck-Condon selection rule
would be lifted.
In this regard, note that breaking of the
Franck-Condon principle occurs
in
multi-level systems such as molecules, where different electronic
transitions intermix because of mutually shared vibrations and
electronic and vibronic transitions are no longer independent.
\cite{X1}

\section{Acknowledgements}
V. V. M. and M. E. R. were supported by the Department of Energy,
Office of Basic Energy Sciences, Grant No. DE-FG02-06ER46313.  

\section{Appendix}

From the definition Eq. (\ref{correlator}) it is apparent that the
$\Delta$-dependence of the correlator is $\cos \Delta T$. Thus, it
is sufficient to carry out calculations for $\Delta=0$. For the
telegraph noise, the correlation function, $K(T)$,
can be expressed as a sum
\begin{eqnarray}\label{Kt}
K(T)={\tilde b}_x^2\sum_{n=0}^\infty(-1)^n \! \!\int\limits_0^\infty\!
\frac{dt_1}\tau e^{-\frac{t_1}\tau} \cdots\!
\int\limits_0^\infty\! \frac{dt_{n+1}}{\tau}
e^{-\frac{t_{n+1}}\tau}
\nonumber\\
\times e^{ib_z(t_1-t_2+\cdots (-1)^nt_{n+1} )}\left[\theta\bigr(\!
{\scriptstyle T-\sum\limits_{k=1}^nt_k }\!\bigl) -\theta\bigr(\!
{\scriptstyle T-\sum\limits_{k=1}^{n+1}t_k }\!\bigl) \right]\!.&&
\end{eqnarray}
Here, $\{t_k\}_{k=1}^{n+1}$ are the time moments at which
$b_z(t)$ flips its sign. The factor $(-1)^n$ ensures that
 $b_x(t)$ flips the sign at the same time moments.
The difference of $\theta$-functions in
Eq. (\ref{Kt}) guarantees that in the time interval $(0,T)$ the
field $b_z$ changes its sign exactly $n$ times, so that
$\sum_{k=1}^nt_k < T < \sum_{k=1}^{n+1}t_k$. Taking the integral
over $t_{n+1}$ by parts leads to
\begin{eqnarray}\label{Kt1}
K(T)={\tilde b}_x^2\sum_{n=0}^\infty(-1)^n \! \!\int\limits_0^\infty\!
\frac{dt_1}\tau \cdots\! \int\limits_0^\infty\! \frac{dt_n}{\tau}
\int\limits_0^\infty\! dt_{n+1} \quad &&
\nonumber\\
\times e^{-\sum_{j=1}^{n+1}\frac{t_j}\tau} e^{ib_z(t_1-t_2+\cdots
(-1)^nt_{n+1} )}\delta\bigr(\! {\scriptstyle
T-\sum\limits_{k=1}^{n+1}t_k }\!\bigl).&&
\end{eqnarray}
Using the integral representation, $\delta(t)
=\int_{-\infty}^\infty \frac{ds}{2\pi}e^{ist}$, in the integrand
and taking the individual integrals yields
\begin{equation}\label{Kt2}
K(T)={\tilde b}_x^2  \sum_{k=1}^\infty~  \!\int\limits_{-\infty}^\infty \!
\frac{ds}{2\pi} e^{isT}\frac {i(s+b_z)\tau^2} {[(1 +is\tau)^2
+b_z^2\tau^2]^k}.
\end{equation}
After taking the sum and symmetrizing with respect to $\pm b_z(0)$,
we find
\begin{equation}\label{Kt3}
K(T)={\tilde b}_x^2\int\limits_{-\infty}^\infty \! \frac{ds}{2\pi}
e^{isT}\frac {is\tau^2} {(1 +is\tau)^2 +b_z^2\tau^2-1}.
\end{equation}
The integral is calculated by finding the contributions of two
poles at $s=\frac{i}{\tau} \pm \left(b_z^2-\frac{1}{\tau^2}\right)^{1/2}$.
For $b_z\tau>1$ we find
\begin{eqnarray}
\label{Ktfin1}
&&K(T)={\tilde b}_x^2\exp{\left(-\frac T\tau\right)}\\
&&\times\left\{\!\cos\! \left[ \frac T\tau
\left(b_z^2\tau^2-1\right)^{1/2} \right] -\frac{\sin\!\left[ \frac
T\tau \left(b_z^2\tau^2-1\right)^{1/2}
\right]}{\left(b_z^2\tau^2-1\right)^{1/2}} \right\}.\nonumber
\end{eqnarray}
For $b_z\tau<1$ Eq. (\ref{Kt3}) yields
\begin{eqnarray}
\label{Ktfin2}
&&K(T)={\tilde b}_x^2\exp{\left(-\frac T\tau\right)}\\
&&\times \left\{\!\cosh\! \left[ \frac T\tau
\left(1-b_z^2\tau^2\right)^{1/2} \right] -\frac{\sinh\!\left[
\frac T\tau \left(1-b_z^2\tau^2\right)^{1/2}
\right]}{\left(1-b_z^2\tau^2\right)^{1/2}} \right\}.\nonumber
\end{eqnarray}

We again emphasize that the factor $(-1)^n$ in the definition Eq.
(\ref{Kt}) had transformed into the signs minus in the final
results Eqs. (\ref{Ktfin1}),  (\ref{Ktfin2}) leading to a peculiar
shape of the two-photon absorption. Physically, repeated changes
of signs of $b_x$ between the same to values\cite{multistep} and
in-phase with $b_z$  leads to averaging out of the effect of
noise.

\end{document}